\documentclass[10pt]{iopart}
\usepackage{graphicx}
\begin{document}

\title
[Electrophoresis of a Rod Macroion under Polyelectrolyte Salt]
{Electrophoresis of a Rod Macroion under Polyelectrolyte Salt:
Is Mobility Reversed for DNA?}

\author{Motohiko Tanaka  
\footnote[3]{Email address: mtanaka@nifs.ac.jp}}

\address{National Institute for Fusion Science, Toki 509-5292, Japan}

\begin{abstract}
By molecular dynamics simulation, we study the charge inversion phenomenon 
of a rod macroion in the presence of polyelectrolyte counterions.  We 
simulate electrophoresis of the macroion under an applied electric field.
When both counterions and coions are polyelectrolytes, charge inversion 
occurs if the line charge density of the counterions is larger than
that of the coions.  For the macroion of surface charge density equal to
that of the DNA, the reversed mobility is realized either with adsorption
of the multivalent counterion polyelectrolyte or the combination of 
electrostatics and other mechanisms including the short-range attraction 
potential or the mechanical twining of polyelectrolyte around the rod axis. 
\end{abstract}




\section{Introduction}

The charge inversion phenomenon takes place due to strong correlations 
of a macroion with small salt ions in solution, which has recently been 
studied for the physiochemical and bio-engineering systems
\cite{Gonza,Walker,Bloomf,Kjell,Netz,Gelbart,NguGS,Mainz,Tanaka,Tanaka2,
Tovar,review,Levin,Tovar2}.
Of particular interest is its possibility of facilitating the delivery of 
genes through negative cell walls \cite{Kabanov,Safin}.
In our previous studies by molecular dynamics simulations 
\cite{Tanaka,Tanaka2}, we adopted both static and dynamic models for the 
macroions.  Specifically, in the dynamical study of electrophoresis with 
explicit (particle) solvent \cite{Tanaka2}, we directly proved the charge 
inversion by measuring the drift of the macroion along the external 
electric field. 
The net charge of the macroion complex was estimated with the use of the 
force balance $ Q^{*} \sim \nu \mu $, where $ \mu $ is the 
electrophoretic mobility and $ \nu $ is the solvent friction.

The purpose of this paper is to demonstrate the possibility of 
mobility reversal under electrophoresis, as the result of charge inversion,
for elongated macroions including the DNA. 
Simulation method and parameters are summarized below.  We take the 
system of one macroion, many counterions, coions and neutral particles
as solvent. 
The units of length, charge and mass are, $ a $, $ e $, and $ m $,
respectively ($ a \sim 1.4 $\AA \ in water and $ m \sim 40 $ a.m.u.)
The rod is assumed to lie perpendicularly to the applied electric 
field which extends fully across the periodic system.
It was shown that a finite-length rod rotates and aligns parallel to 
the electric field due to the rod polarization by specific condensation 
of positive ions at one end toward the electric field and the negative 
ions at the other end \cite{Netz,Tanaka4}. 
Charge inversion occurs similarly in that case, but the discreteness 
of the surface charge is essential since the counterions need to be 
pinned down on the smooth surface of the rod macroion \cite{Tanaka4}. 

The macroion has a radius $ R_{0}= 5a $, negative charge $ Q_{0} $
with $ -20e $ or $ -80e $, and mass $ 2000m $, which is surrounded 
by the $ N^{+} $ number of counterions of a positive charge $ Z^{+}e $
and the $ N^{-} $ coions of a negative charge $ -Z^{-}e $.
The system is maintained in overall charge neutrality, $ Q_{0} +
N^{+}Z^{+}e - N^{-}Z^{-}e = 0 $.
The radii of counterions and coions are $ a^{+} $ and $ a^{-} $,
respectively, with the counterion radius being fixed $ a^{+}= a $,
and that of neutral particles is $ a/2 $.  The mass of the coions and
counterions is $ m $, and that of $ N_{*} $ neutral particles is
$ m/2 $. 
Approximately one neutral particle is distributed in every volume
element $ (2.1a)^{3} \approx (3 $\AA$)^{3} $ inside the periodic
simulation domain of the side $ L=32a \approx 45 $\AA, excluding 
the locations already occupied by ions. 

We solve the Newton equations of motion for each particle with the 
electrostatic (Coulombic) and Lennard-Jones potential forces under 
the uniform applied electric field $ E $ ($ E > 0 $).
A large number of neutral particles are used to model the viscous
solvent of given temperature and to treat the interactions among the
finite-size macroion, counterions, coions and solvent.
The Coulombic forces under the periodic boundary conditions \cite{Ewald}
are calculated efficiently with the use of the so-called 
particle-particle-particle-mesh algorithm \cite{Eastwood,Deserno},
with $ (32)^{3} $ spatial meshes, the real-space cutoff $ 10a $ and the 
Ewald parameter $ \alpha \approx 0.262 $ (the electric field is 
accurately calculated with the Ewald method).  
The volume exclusion effects between particles (both charged and
non-charged) are treated with the repulsive Lennard-Jones potential 
$ \phi_{LJ}= 4 \varepsilon [(A/r_{ij} )^{12} - ( A/r_{ij} )^{6}] $ 
for $ r_{ij}= |{\bf r}_{i}-{\bf r}_{j}| \le 2^{1/6}A $, and 
$ \phi_{LJ}= - \varepsilon $ otherwise. 
Here $ {\bf r}_{i}$ is the position vector of the $ i $-th particle,
and $ A $ is the sum of the radii of two interacting particles.
We relate $ \varepsilon $ with the temperature by
$ \varepsilon = k_{B}T $, and choose $ k_{B}T = e^{2}/5 \epsilon a $
(we assume spatially homogeneous dielectric constant $ \epsilon $).
The Bjerrum length is thus $ \lambda_{B}= e^{2}/\epsilon k_{B}T=
5a $, which is 7 \AA \ in water.
The equations of motion are integrated with the use of the
leapfrog method, which is equivalent to the Verlet algorithm \cite{textMD}.
The unit of time is $ \tau = a \sqrt{m/ \varepsilon} $ ($ \approx $ 1ps),
and we choose the integration time step $ \Delta t= 0.01 \tau $.

The Joule heat produced by the external electric field on ions
and transferred to neutral particles by collisions is drained
by a heat bath for neutral particles at the boundaries.
The thermal bath which screens hydrodynamic interactions is safely
adopted in the present study since the hydrodynamic interactions 
are screened at short distances in the electrolyte solvent 
\cite{Ajdari,Viovy}. We confirmed this fact by good agreement of 
our results with and without the heat bath \cite{Tanaka2}.
We also note that the effects of finite length and rotation 
of the rod macroion were discussed in \cite{Tanaka4,Tanaka3}.

\section{A Rod Macroion with Polyelectrolytes}
\subsection{Polyelectrolyte Counterions and Coions}

We showed previously that the mobility of a rod (cylindrical) macroion  
reversed its sign at zero salt, which was enhanced by addition of small 
amount of monovalent salt if the macroion was strongly charged, 
$ \sigma_{rod} = |Q_{rod}|/2 \pi R_{rod}L > 0.04e/a^{2} $ 
(0.33 C/m$ {}^{2}$) \cite{Tanaka3}. 
For reference, the surface charge density of the DNA is 0.19 C/m$ {}^{2} $.
The reversed mobility of the rod macroion was more persistent to larger
amount of monovalent salt than the spherical macroion of the same radius 
and surface charge density. However, the rod macroion with surface charge 
density equal to that of the DNA was not subject to mobility reversal.  
As will be shown in this paper, polyelectrolyte counterions can promote 
overcharging of the macroion \cite{Silva}.

\begin{figure}
\centerline{\resizebox{0.98\textwidth}{!}{\includegraphics{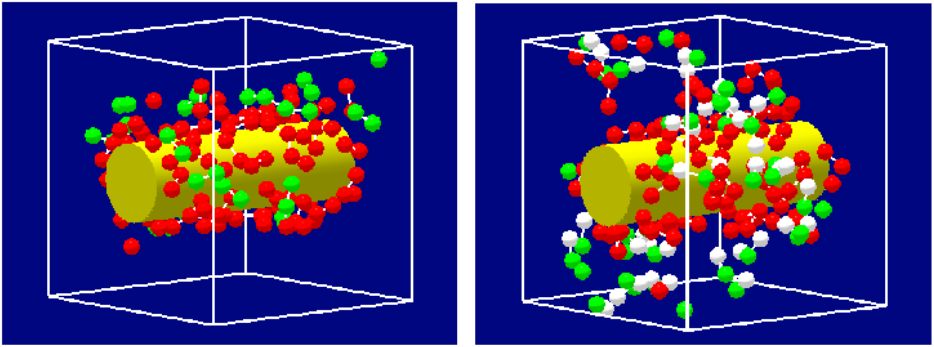}}}

\vspace*{-0.35cm}
\caption{The rod-shaped macroion under polyelectrolyte counterions and coions.
All chains consist of four monomers; each of 30 counterion polyelectrolyte chains 
carries four unit charges $ e $ (red monomers). 
The coion polyelectrolyte is either (left) 10 chains consisting of four unit 
charges $ (-e) $ (green monomers), or (right) 20 chains of two unit charges 
and two neutral monomers (white monomers).}

\label{bird-polyel} 
\end{figure}

Figure \ref{bird-polyel} shows the charge inversion of a rod macroion when both 
the counterions and coions are polyelectrolytes. For simplicity, all chains have
the same length of four monomers. 
All the monomers are charged unity $ e $ for 30 chains of counterion 
polyelectrolyte (red monomers). 
For the coions, there are two cases: (a) 10 chains of four unit charges $ (-e) $ 
(green monomers), and (b) 20 chains, each consisting of two unit charges $ (-e) $ 
and two neutral monomers (white).  
When the line charge densities of the counterion and coion polyelectrolytes are 
equal as in Fig.\ref{bird-polyel}(a), the mobility reversal is small 
as shown in Table I (the bin of $ -4e \times 10 $ and "no salt"). 
This is because the well adsorbed coion polyelectrolyte cancel charges of the 
counterion polyelectrolyte.
When the line charge density of the coion polyelectrolyte is reduced to a half, 
adsorption of coion polyelectrolyte becomes less as shown in 
Fig.\ref{bird-polyel}(b), and significant revered mobility is obtained (the bin 
of $ -2e \times 20 $ and "no salt").
Further reduction of the line charge density of coion polyelectrolyte yields 
nearly the same reversed mobility.  
When the monovalent salt whose charge content is twice as large as that of 
the coion polyelectrolyte is added, the mobility of the macroion 
is non-reversed (the bottom row of Table I). 
Charge neutralization by the coion polyelectrolyte is good when its line
charge density is comparable to that of the counterion polyelectrolyte. 

\begin{table}
\caption{ The mobility of the infinite rod macroion $ Q_{rod}= -80e $
under polyelectrolyte counterions and coions.  Among 30 counterion
polyelectrolyte chains consisting of four unit charges $ e $, 20 chains
are required to neutralize the macroion.
The line charge density of the coion polyelectrolyte is varied, where each 
chain consists of one, two or four unit charges $ (-e) $ and interleaving
neutral monomers for the rest.
Also, the case with monovalent salt, $ N_{salt}^{+}= 80 $ and
$ N_{salt}^{-}=80 $ is shown in the bottom row.
The mobility values are in unit of $ 10^{-2} \mu_{0r} $.}

\vspace*{0.2cm}
\begin{indented}
\item[]
\begin{tabular}{ccccc} \hline
charges $\times$ chains & \vline & $ -4e \times 10 $ & $ -2e \times 20 $ & $ -e \times 40 $  \\
\hline
no salt      & \vline & $  1.3 $ & $ 3.2   $ & $ 3.4 $  \\
w/ salt      & \vline & $ -4.9 $ & $-0.085 $ & $-0.36 $ \\
\hline
\end{tabular}
\end{indented}
\end{table}

\subsection{A Rod Macroion with Surface Charge Density Equal to the DNA}

Previous studies of charge-inverted rod macroions (polyelectrolyte)
dealt with their condensation to a charged surface \cite{Netz,Gelbart,NguGS}, 
or the adsorption of metal ions on themselves \cite{Gonza} except for a 
very recent study \cite{Messina2}. Here, we treat 
the charge inversion of a cylindrical macroion in the presence of
polyelectrolyte counterions, where the surface charge density of the macroion 
is approximately that of the DNA, $ \sigma_{rod} \approx 0.02e/a^{2} $. 
The macroion is assumed to be an infinite rod with $ Q_{rod}= -20e $, radius 
$ R_{rod}= 5a $ and mass $ 2000m $, lying perpendicularly to the applied 
electric field. The surface charges are provided by two sets of ten discrete 
charges $ (-e) $ aligned helically at the depth $ a $ below the surface of 
the rod; the spacing of unit charges is $ 4.1a $ along the helix contour. 
The polymer counterions (polyelectrolyte) consist of the monomers either 
of (i) trivalent ions or (ii) monovalent ions.  
The coions are monovalent with larger radius than the counterions 
$ a^{-}/a^{+}= 1.5 $, which is favorable for charge inversion \cite{Tanaka3}.
The Bjerrum length is $ e^{2}/\epsilon k_{B}T= 5a $. 
The mobility normalization is $ \mu_{0r}= v_{0}/(2|Q_{rod}|/R_{rod}L) =
v_{0}/4 \pi \sigma_{rod} \approx 2.8 \times 10^{-4} $(cm/s)/(V/cm), 
where $ v_{0} $ is the thermal speed of neutral particles.
The Z-ion concentration at $ n_{zI} \approx 0.005/a^{3} $ is approximately 
0.33 M/l.

\begin{figure}
\centerline{\resizebox{0.60\textwidth}{!}{\includegraphics{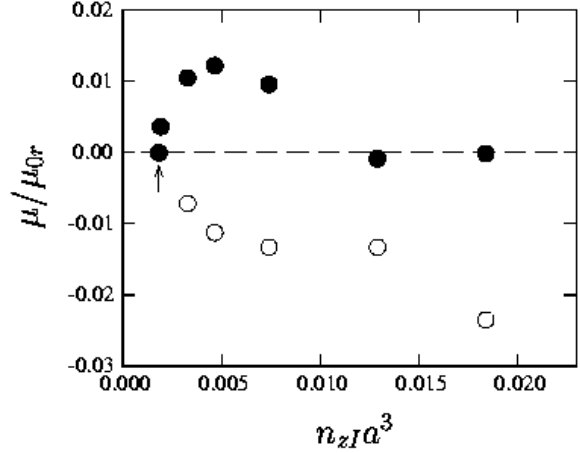}}}

\vspace*{-0.3cm}
\caption{ The mobility of a cylindrical macroion against the ionic 
strength of polyelectrolyte counterions $ n_{zI}= Z^{2}N^{+}/L^{3} $, 
where $ \mu_{0r} \approx 2.8 \times 10^{-4} $ (cm/s)/(V/cm) and 
$ 0.005e/a^{3} \approx 0.33 $ M/l.  
Unit charges placed along double helices provide the macroion surface 
charge with $ \sigma_{rod} \approx 0.02e/a^{2} $ (0.17 C/m$ {}^{2}$ 
$ \sim \sigma_{DNA} $).
(a) Each counterion chain consists of three trivalent Z-ions (filled 
circles), (b) the same as (a) except that all counterions are isolated 
spheres (open circles).  
The coions are larger than the counterions $ a^{-}/a^{+}= 
1.5 $.  The arrow shows the isoelectric point.}
\label{Fig.mu-DNA}
\end{figure}

\begin{figure}
\centerline{\resizebox{0.98\textwidth}{!}{\includegraphics{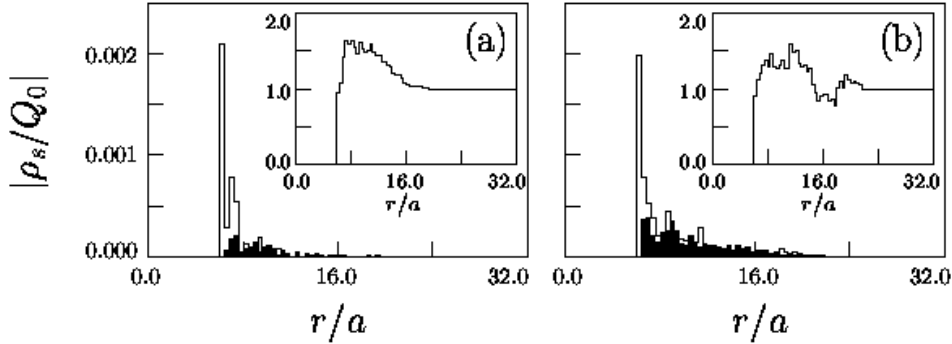}}}

\vspace*{-0.3cm}
\caption{The distribution functions of the counterions (open bars) and
coions (shaded bars) for the rod macroion with double helix charges whose 
surface charge is equal to that of the DNA. These correspond to the 
filled circles in Fig.\ref{Fig.mu-DNA} with the ionic strength 
(a) $ n_{zI} \sim 0.005a^{-3} $ and (b) $ n_{zI} \sim 0.013a^{-3} $.}
\label{Fig.distr-DNA} 
\end{figure}

\begin{figure}
\centerline{\resizebox{0.98\textwidth}{!}{\includegraphics{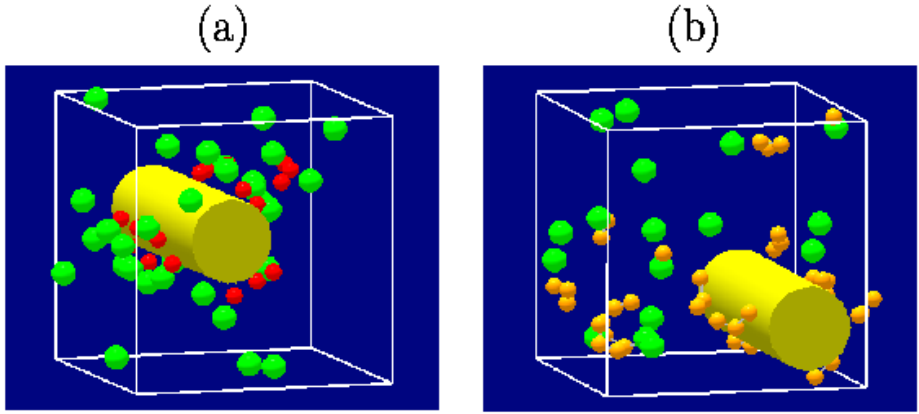}}}

\vspace*{-0.3cm}
\caption{A rod macroion whose surface charges are provided by discrete
unit charges along double helices with the surface charge density nearly 
equal to that of the DNA, is shown for (a) polyelectrolyte cations 
of three trivalent ions $ N^{+3}=17 $ (red spheres), and (b) 
polyelectrolyte of three monovalent ions $ N^{+}=35 $ (yellow spheres).   
The former corresponds to the filled circle in Fig.\ref{Fig.mu-DNA}
at $ n_{zI} \sim 0.005/a^{3} $. 
The anions (green spheres) are larger than the cations, $ a^{-}/a^{+}= 1.5 $.  
The external electric field points horizontally rightward.}
\label{Fig.bird-DNA} 
\end{figure}

For the counterion polyelectrolyte with $ 3e-3e-3e $ monomers, the macroion
mobility is reversed as denoted by filled circles in Fig.\ref{Fig.mu-DNA}. 
The reversed mobility increases with the ionic strength of 
Z-ions as $ n_{zI}^{1/2} $ above the isoelectric point (arrow), 
as predicted by theory \cite{NguGS}.  Then, the mobility
decreases and gets non-reversed since more coions condense on 
the counterion polyelectrolyte as the number of screening coions 
increases with the ionic strength.
This is verified in the radial distribution functions in 
Fig.\ref{Fig.distr-DNA}, where the profile of the integrated
charge for the large ionic strength in Fig.\ref{Fig.distr-DNA}(b)
has an overshoot and is less prominent. 
This overshoot is characteristic of non-reversed
mobility in spite of the peaked charge profile \cite{Tanaka3}. 
At the point where the reversed mobility terminates, the charge 
content of the salt in the surface layer is still less than that 
of the macroion surface charge, $  Ze c_{s}\lambda_{D}  
\sim \frac{1}{4} \sigma_{rod} $.
Good adsorption of the polyelectrolyte Z-ions is seen in the 
bird's-eye view plot of Fig.\ref{Fig.bird-DNA}(a), where the average 
bond length between the chained Z-ions is $ 3.2a $.
This behavior agrees with the result of the theory \cite{Toan3}.
Also, it may be related to the electrophoresis experiment that observed
reversed mobility for the concentration ratio of [polycation]/[DNA] $ 
\approx 2 $, where the macroion was a chicken embryo DNA and the 
polycation was monovalent polymer with a long hydrophobic chain 
$ (CH_{2})_{n}CH(C_{5}NH_{4})^{+}C_{2}H_{5} $ \cite{DNA-ep}.
The differences of the experiment from the simulation setting, however, 
are the monovalence and existence of the hydrophobic chain,
which will be mentioned in the next subsection.

For isolated Z-ions with other conditions fixed as above, the mobility
reversal does not occur (open circles in Fig.\ref{Fig.mu-DNA}).  The
counterion condensation is weak in this case, possibly due to desorption 
of counterions by thermal agitation.  The correlation energy 
between the surface Z-ions, calculated with the Wigner-Seitz cell 
radius $ R_{WS}= (Z/\pi \sigma_{rod})^{1/2} \sim 6.9a $, 
\begin{eqnarray}
 Z^{2}e^{2}/2 \epsilon R_{WS} \sim 3.3 k_{B}T   
\end{eqnarray}
is comparable to thermal energy, which is below the threshold of
charge inversion.
Thus, the polyelectrolyte counterions with a rod macroion is more 
favorable for the mobility reversal than with the isolated counterions 
for the same ionic strength \cite{Tanaka4}.  

For a strongly charged rod macroion with $ \sigma_{rod} \sim 0.08e/a^{2} $, 
the mobility is reversed and increases above the isoelectric point. 
The peak mobility occurs around $ n_{zI} \sim 0.02a^{-3} $ with $ \mu 
\sim 0.04 \mu_{0r} $. 
The Z-ion correlation energy here is twice large compared to the previous 
case, $ Z^{2}e^{2}/2 \epsilon R_{WS} \sim 6.5 k_{B}T $ ($ R_{WS} \sim 3.5a $).  
The reversed mobility, i.e. net overcharging, increases in proportion to 
the macroion bare charge.
The theory of an infinite-length rod adsorption on a two dimensional plane 
can be applied to the present simulation if one makes the following
replacements, the DNA $ \rightarrow $ polyelectrolyte, the planar surface
$ \rightarrow $ the rod macroion, which predicts net overcharging 
\cite{Netz,NguGS} 
\begin{eqnarray}
Q^{*} \sim (\eta/\lambda_{D})/ \ln(\eta/\sigma_{rod}\lambda_{D}).
\end{eqnarray}
Here, the line charge density of the polyelectrolyte $ \eta \sim e/3.2a $ 
predicts 2.6 times more overcharging for the strongly charged macroion 
than the weakly charged macroion.  Our result agrees well with the
theory prediction.

As the second case, the electrophoresis of the same macroion under all 
unit-charge polyelectrolyte is examined.  
The short polyelectrolyte whose charge content (chain length) 
is less than $ 10e $ is not subject to charge inversion.  
The electrostatic energy between the unit charges with the mutual
distance $ b \sim 3a $ is $ e^{2}/\epsilon b \sim \frac{5}{3}
k_{B}T $.  Although the Manning-Onsager condensation of coions 
\cite{Manning} on the counterion polyelectrolyte is not evident in 
Fig.\ref{Fig.bird-DNA}(b), the adsorption of such polyelectrolyte 
is not as strong as to cause charge inversion.  The radial distribution 
function and the bird's-eye view plot of Fig.\ref{Fig.bird-DNA}(b) 
reveal that about a half of such polyelectrolyte chains are located 
away from the macroion surface, thus the net charge of the macroion 
complex is negative.  
It is mentioned in passing that the present molecular dynamics 
simulation is not reproducing the charge inversion due to optimal 
adsorption of unit-charge polyelectrolyte through fractionalization 
\cite{Toan2}.  

\begin{figure}
\centerline{\resizebox{0.47\textwidth}{!}{\includegraphics{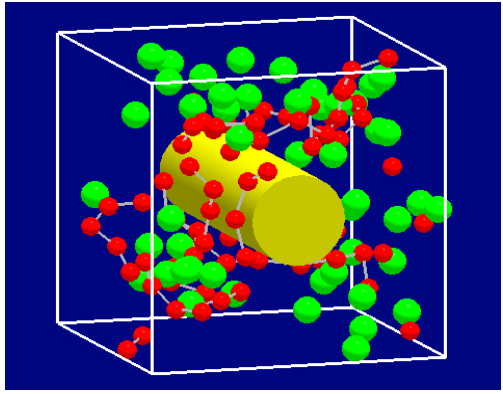}}}

\vspace*{0.3cm} 
\caption{A cylindrical macroion with long polyelectrolyte cations
of unit charges (red) and isolated anions (green) has reversed mobility.   
The surface charge density of the macroion with double helix charges, 
$ \sigma_{rod} \sim 0.02e/a^{2} $, is approximately that of the DNA. 
The long chains of polyelectrolyte mechanically wind around the rod. 
The radius and number of anions are $ a^{-}= 1.5a $ and $ N^{-}= 60 $, 
respectively [neutral particles are not shown].}
\label{Fig.birdeye-DNA}
\end{figure}

\subsection{Non-Electrostatic Mechanisms for Mobility Reversal}

As mentioned above, the unit-charge polyelectrolyte of reasonable
length alone does not lead to mobility reversal.
Nevertheless, we show below that the mobility reversal occurs for such 
polyelectrolytes if another mechanism is superimposed. For example,
the polymers with hydrophobic tails feel short-range attraction 
forces around the DNA. Here, this effect is modeled with the 
attractive Lennard-Jones potential of the depth $ 1 k_{B}T $, without 
truncation of the potential at the distance $ 2^{1/6} \sigma $. 
A reversed mobility is obtained for the polyelectrolyte 
counterions of 10 unit charges, $ \mu \approx 0.015 \mu_{0r} $.
An analytical theory showed reversed mobility for the surfactant 
having a charged head group and hydrophobic tails with strong 
hydrophobicity $ \chi= -6k_{B}T $ \cite{Silva}.
Our molecular dynamics simulation uses a weak attraction potential
but requires rather a long chain.  More works are necessary to
reconcile these results with the experiment using the DNA as the
macroion and monovalent counterion polyelectrolyte \cite{DNA-ep}.

The following simulation result gives an implication where large 
reversed mobility $ \mu \approx 0.073 \mu_{0} $ occurs for a very 
long polyelectrolyte chain of $ 20e $. 
This mobility reversal is due to mechanical twining of the 
polyelectrolyte chain around the rod axis, as depicted in 
Fig.\ref{Fig.birdeye-DNA}.  
These chains are not statically attached to the macroion, but they 
pull the macroion by twining toward the electric field on time average.  
However, when the chain length is cut to a half with the total number of 
counterions fixed, mobility is not reversed as the chains can pass 
by the rod.
Although the mechanical twining of polyelectrolyte counterions is 
not due to electrostatics, both the electrostatic and non-electrostatic 
mechanisms can work together to cause the mobility reversal of the DNA 
macroion in the electrophoresis experiments.

\section{Summary}

In this paper, the charge inversion phenomenon of a rod macroion 
under polyelectrolyte counterions was studied. 
When both the counterions and coions were polyelectrolytes, the 
charge inversion took place if the line charge density of the 
counterions prevailed over that of coions.
The mobility was reversed for the rod macroion of the surface charge 
density equal to that of the DNA due to polyelectrolyte counterions
of multi-valence.

For short chains of counterion polyelectrolyte consisting of unit 
charges, the electrostatic effect alone was not sufficient to cause 
mobility reversal.  
Either the short-range attraction due to hydrophobicity or the 
mechanical twining of the polyelectrolyte around the rod macroion 
could cause the mobility reversal.  
In real environments, the electrostatic mechanism and other 
effects such as hydrophobic attraction by specific configurations 
or mechanical twining of the polyelectrolytes may collaborate to 
result in the mobility reversal under the electrophoresis.

\vspace*{0.5cm} 
\noindent{\bf Acknowledgments}

\vspace*{0.2cm} 
The author cordially thanks Prof.A.Yu.Grosberg for close collaborations 
on the molecular dynamics study of the charge inversion phenomenon.  
This study and the travel to participate in {\it The International 
Conference of Applied Statistical Physics of Molecular Engineering 
2003} (Mexico) was supported by Grant in Aid No.15035218 (2003) 
from the Ministry of Education, Science and Culture of Japan. 
The computation was performed with the computers of 
the University of Minnesota Supercomputing Institute, and those 
of the Institute for Space and Astronautical Science of Japan.

\end{document}